# A study of research trends and issues in wireless ad hoc networks

**Noman Islam**
*Technology Promotion International, Karachi, Pakistan*
**Zubair A. Shaikh**
*National University of Computer and Emerging Sciences, Karachi, Pakistan*

## ABSTRACT

Ad hoc network enables network creation on the fly without support of any predefined infrastructure. The spontaneous erection of networks in anytime and anywhere fashion enables development of various novel applications based on ad hoc networks. However, at the same ad hoc network presents several new challenges. Different research proposals have came forward to resolve these challenges. This chapter provides a survey of current issues, solutions and research trends in wireless ad hoc network. Even though various surveys are already available on the topic, rapid developments in recent years call for an updated account on this topic. The chapter has been organized as follows. In the first part of the chapter, various ad hoc network's issues arising at different layers of TCP/IP protocol stack are presented. An overview of research proposals to address each of these issues is also provided. The second part of the chapter investigates various emerging models of ad hoc networks, discusses their distinctive properties and highlights various research issues arising due to these properties. We specifically provide discussion on ad hoc grids, ad hoc clouds, wireless mesh networks and cognitive radio ad hoc networks. The chapter ends with presenting summary of the current research on ad hoc network, ignored research areas and directions for further research.

## INTRODUCTION

During last few years, extensive developments have been observed in the domain of wireless network. Different communication technologies i.e. general packet radio service (GPRS), enhanced data rates for GSM evolution (EDGE) and worldwide interoperability for microwave access (WIMAX) etc. have evolved and newer form of computing devices i.e. personal digital assistant (PDA), tablets and smart phones are appearing in the market. The wireless computing has progressed from 1G to 4G communication networks. During this progression, various modes of wireless networking have emerged. The simplest form of wireless networking is communication among two or more fixed hosts in open air. The conventional television system operates on this mode. Another approach is *wireless networking with access point*. There are different wireless hosts that are allowed to move while the basic infrastructure is supported by set of fixed nodes called base stations or access points. However, this approach doesn't provide the flexibility to be used in emergency situations requiring quick deployment or networking in adversarial surroundings. The evolution of technologies has lead to development a new mode of wireless networking where the nodes arrange themselves on the fly in the form of a network without any infrastructure support. Such networks are called ad hoc networks.

## Properties of ad hoc network

Formally, Ad hoc Network $G(N,E)$ is defined as a collection of nodes $N=\{n_1,n_2,n_3,\ldots\}$ connected by edges $E \subseteq N \times N$ (Islam and Shaikh 2012). The nodes are usually mobile with limited capabilities, links are volatile and insecure, and there are no dedicated nodes for addressing, routing, key management and directory maintenance etc. The nodes are themselves responsible for various network operations i.e. routing, security, addressing and key management etc. It is obvious from these characteristics that network protocols and algorithm that are currently available for wired and infrastructure-less wireless networks are not suitable for ad hoc networks (Islam, Shaikh et al. 2010). For example, a conventional routing algorithm when employed for ad hoc network can suffer from loops, stale routes and other issues due to the very sharp changes in the network. Similarly, the current security solutions are based on availability of authentication servers, certification authority and other security infrastructure, which are not generally available in ad hoc network. Therefore, new solutions are required for addressing various challenges of ad hoc network.

Different research efforts are underway to address various issues of ad hoc networks. In this chapter, we provide an adequate account of these efforts. There are already some surveys available that have summarized the previous researches on ad hoc networks. For example, Dow, Lin et al. (2005) and Singh, Dutta et al. (2012) have provided a quantitative analysis of the number of research proposals appeared during last few years for addressing a particular issue of ad hoc network. Similarly, a summary of various research issues in ad hoc networks have been presented in (Chlamtac, Conti et al. 2003; Toh, Mahonen et al. 2005; Ghosekar, Katkar et al. 2010; K.Al-Omari and Sumari 2010). However, the focus of this chapter is on research pursued in ad hoc networks during recent years. The major contributions of this chapter are as follows:

- To provide a summary of various research issues in ad hoc networks and the recent approaches adopted to tackle these issues
- To investigate and report on various emerging models of ad hoc networking
- To present a comprehensive overview of issues and corresponding solutions for different ad hoc networking models i.e. ad hoc grids, ad hoc clouds, wireless mesh networks and cognitive radio ad hoc networks etc.
- To summarize the current state-of-the-art and avenues for further research

## Research issues in ad hoc network

We start the discussion with an overview of major research issues in ad hoc networks. As discussed earlier, the issues arising in ad hoc network span across all layers of communication. In addition, cross-layer issues i.e. security, quality of service (QoS) and energy management ad hoc networks etc. demand resolution mechanisms at more than one layer of communication. Different research efforts have been put in to address these issues. Let's discuss the research issues of ad hoc network along with the proposed mechanisms to address them.

### Antenna Design

Ad hoc networks are characterized by limited channel capacity and unreliable links. In addition there are presence of noise and interference in the surrounding environment. Therefore, new antenna design techniques are required that can cope with these limitations. One of the approaches to address these problems is to have multiple antenna elements to improve the quality of received signals (Ramanathan 2001). In *switched diversity antenna* design, the antenna element is changed continuously and the system uses the element with best gain. This solves the problem of fading and multipath effects. Similarly, *diversity combining* is based on correcting the phase error and combining the power to have more gain. To improve the spectral efficiency, directional antenna has also been proposed in recent literature. This approach leads to efficient utilization of spectrum and more power by pointing the signal in a specified direction. Two types of approaches can be used for directional transmission (Ramanathan, Redi et al. 2005). In a *switched beam antenna*, there are fixed beams that can be formed by shifting the phase of antenna by a fixed amount or by switching between various fixed directional antennas. An *steered antenna* can be focused in any direction.

The use of smart antennas for ad hoc networks, calls for addressing multitude of research questions. For instance, at medium access control (MAC) layer, various issues can arise (Bazan and Jaseemuddin 2012). A node while transmitting is deaf except the direction in which it is transmitting. Hence, it can't respond to RTS/CTS messages. Similarly, head of line blocking occurs, when a node finds the medium busy in the direction in which it is transmitting. It holds the other messages in its queue, even though other messages can be transmitted in a different direction. Various specialized MAC layer protocols have thus been proposed for smart antenna based systems and can be seen in (Bazan and Jaseemuddin 2012; Lu, Towsley et al. 2012). Another important question pertains to size and cost of smart antenna. This is dependent on the domain of application. For example, a military network can afford the cost of smart antenna deployments as well as the size of antenna doesn't matter in such applications. However, for small scale applications, further research is required to design cost efficient, small sized antennas.

### Energy Management

In ad hoc networks, the nodes have very limited battery power. Algorithms for ad hoc networks should be designed such that they consume minimum energy for their execution. Substantial efforts have been put in to devise energy efficient algorithms for ad hoc networks. One of the techniques is to *control the transmission power* of the nodes. This is done by adjusting the transmission power to an arbitrary value to obtain optimal interference, power savings and improved channel capacity. A good survey on various power control algorithms is provided in (Singh and Kumar 2010). However, the connectivity of the nodes is affected by the transmission power and requires careful design. Gomez and Campbeel (2007) have analyzed the impact of power control on connectivity, power savings and capacity of the network.

Another approach to energy management is to devise algorithms such that nodes remain in *low energy conservation state* most of the time. A wireless node can be in transmission, receiving, idle listening or sleep state at particular instant of time. In the sleep state, the transceiver of the node is off and consumes a very small amount of energy i.e. 0.740 W (Fotino and De Rango 2011). Hence, an energy efficient algorithm should keep the nodes in sleeping mode most of the time. There have been various algorithms proposed in literature for power management based on smart utilization of sleep states. Lim, Yu et al. (2009) have proposed random cast, an energy efficient scheme over dynamic source routing (DSR) protocol based on IEEE 802.11 power saving mechanisms.

Another approach to energy management is energy efficient routing. An *energy efficient routing algorithm* employs a routing metric that reflects the energy consumption during routing operation. The metrics that can be used are residual battery of the nodes and energy consumed to forward the packet along the route ((Hassanein 2006; Kwon and Shroff 2006). There are also some routing protocols that attempt to organize the networks in to clusters such that the number of message exchanges can be minimized (Heinzelman, Chandrakasan et al. 2000; Tariquea, Vitae et al. 2009). This leads to reduced power consumption by nodes. Finally, there are *multipath routing protocols* that distribute the overall routing process across multiple paths and reduce the power consumptions at a particular node (Shah and Rabaey 2002; Liu, Guo et al. 2009).

### MAC Layer protocols

A MAC layer protocol provides a fair access to the medium for transmission of information. The various MAC schemes proposed in literature are classified as contention free schemes and contention based schemes (Kumar, Raghavan et al. 2006). The *contention free scheme* is based on assignment of time and frequency, and is free from collisions. Examples are Frequency Division Multiple Access (FDMA), Time Division Multiple Access (TDMA) and Code Division Multiple Access (CDMA) protocols etc.

In the *contention base schemes*, the nodes compete for access to the medium. Carrier Sense Multiple Access (CSMA) is the classical example of contention based protocols. In CSMA, the node first senses the medium before any transmission. The node can defer its transmission in case any other message is being transmitted. CSMA protocols however suffer from hidden node and exposed node problems (Kumar, Raghavan et al. 2006). Multiple Access Collision Avoidance (MACA) protocol attempts to solve this problem by transmission of ready to send (RTS) and clear to send (CTS) messages before commencing the transmission. The nodes hearing the RTS messages defer their transmission. MACA-By Invitation (MACA-BI) is a receiver initiated protocol where the receiver requests the sender for transmission by sending ready to receive (RTR) messages (Talucc 1997). Thus the sequence of message exchanges is RTR-DATA.

IEEE also specifies the distributed coordination function (DCF) and point coordination function (PCF) as access control protocols for wireless network. They are based on RTS-CTS-DATA-ACK sequence of messages for data transmission. A Network Allocation Vector (NAV) is also maintained that represents the expected duration during which the wireless medium will be busy. Nodes update NAV values based on overhearing the transmissions. More details can be seen in (Chen 1994).

There has been lot of research work done on designing *QoS aware MAC protocols*. Shin, Yun et al. (2011) proposed a multichannel MAC protocol that enables QoS over IEEE 802.11 DCF. Kamruzzaman, Hamdi et al. (2010) proposed a QoS based MAC protocol for cognitive radio ad hoc networks. In addition, there have been some specialized *MAC protocols for directional antennas* (Wang, Zhai et al. 2008). As the directional antenna transmits in a particular direction, these specialized MAC protocols make optimum use of spectrum as well as address the challenges arise due to the directional antennas, as discussed in previous section.

### Routing

One of the features of ad hoc network is multi-hopped routing in which every node has to participate in the routing operation. This presents a number of challenges i.e. creation of false routes, routing loops and security attacks on routing protocols etc. Routing protocols proposed for ad hoc networks can be classified as proactive, reactive and hybrid routing protocols. In a *proactive* routing protocol also known as table drive routing protocol, every node periodically disseminate its routing updates to other nodes. Destination Sequenced Distance Vector (DSDV), Global State Routing protocol (GSR), Fisheye State Routing (FSR), Optimized Link State Routing (OLSR) and Cluster Gateway Switch Routing Protocol (CGSR) are examples of some of the proactive routing protocols (Kumar, Reddyr et al. 2010). *DSDV* is based on periodic / event-based dissemination of routes to neighbors. A sequence number is maintained to determine freshness of the routing entry and avoid inconsistencies in routing information. *GSR* is based on exchange of link state information in the form of vectors with neighboring nodes. Based on the information exchanges, global topology of the network is constructed at each node. In *FSR*, the nodes prioritize other nodes based on their distances with other nodes and the routing updates are exchanged with nearer nodes more frequently than distant nodes, thus reducing congestion in network. *OLSR* is an optimization of the link state routing used in wired networks. *CGSR* is a hierarchical protocol based on DSDV that divides the whole networks in to clusters such that routing overhead can be minimized.

The second type of routing protocol presented in literature is *reactive* routing protocol also known as on-demand routing protocol. The routing operation is performed on as need basis. Among the various reactive protocols, the popular ones are Ad hoc On-Demand Distance Vector Routing (AODV), Distance Vector Routing (DSR), Temporally Ordered Routing (TORA) and Cluster Based Routing (CBR) protocols. The *AODV* and *DSR* are based on the broadcasting of a route discovery request (RREQ) to neighbors until the request reaches the destination. The destination sends a reply which reaches back to the source using the same route that is used for request's dissemination. In the process of dissemination, a route is created towards the source. *TORA* is a reactive routing protocol that creates a route towards the destination based on a directed acyclic graph from source to destination. *CBR* protocol is a hierarchical protocol based on formation of clusters among the network nodes, thus minimizing the messages exchanged in the network. The combination of proactive and reactive protocols i.e. *hybrid routing protocols* have also been advocated in some research proposals. Among them are *Zone Routing Protocol (ZRP)*, *Dual-Hybrid Adaptive Routing (DHAR), Adaptive Distance Vector Routing (ADV)* and *Sharp Hybrid Adaptive Routing Protocol (SHARP)* etc. (Kumar, Reddyr et al. 2010).

Some routing protocols exploit position information for routing of packets. They are called *position based routing protocols*. There are three popular approaches to position based routing in ad hoc networks. These are *greedy forwarding*, *restricted direction flooding* and *hierarchical approaches* (Qabajeh, Kiah et al. 2009). A greedy forwarding routing protocol doesn't maintain the complete route towards the destination. During forwarding, the packet includes the geographical location of the recipient node. The intermediate nodes propagate the request towards the destination based on some optimization criteria like the next neighbor closest to the destination. An example of such scheme is Most Forward within distance R (MFR) (Takagi and Kleinrock 1983). In this protocol, a packet is forwarded to the next hop node which has the maximum progress towards the source. A projection line (with respect to next hop node) is defined as the line between sender and receiver and the progress is the distance from the sender to next hop on the projection line. Improved progress Position Based Beacon Less Routing algorithm (I-PBBLR) extends this approach further by incorporating the direction metric to improve the progress definition (Cao and Xie 2005). In *restricted direction flooding* protocols, the routing request is propagated towards the destination only in forward direction. The receiving nodes check if they are set of nodes towards the destination and if they should forward the packet ahead. In that case, the nodes retransmit the packet. An example of this type of protocol is Distance Routing Effect Algorithm for Mobility (DREAM) that propagates the routing request in a particular sector (Qabajeh, Kiah et al. 2009). The third strategy is based on forming a hierarchy among the nodes based on location and other parameters. An example of this approach is TERMINODES that maintains a two level hierarchy among the nodes (Qabajeh, Kiah et al. 2009). If the location of destination node is closer to the source, packets will be routed based on a proactive routing protocol. Greedy routing will be used for routing to distant nodes.

In addition, there are QoS based routing protocols that selects a route based on QoS attributes of the links. Various QoS routing protocols have emerged during recent years. *Core-Extraction Distributed Ad hoc Routing (CEDAR)* is based on establishing a core structure that provides a low over ahead infrastructure for QoS routing (Sivakumar, Sinha et al. 1999). As the infrastructure comprises stable nodes, the desired QoS of user can be easily maintained. In *Multipath Routing protocol (MRP)*, several alternate paths towards the destination are discovered based on bandwidth and reliability constraints (Qin and Liu 2009). Several *extensions over traditional routing protocols* are also proposed to cope with QoS requirements (Gangwar, Pal et al. 2012). For example, *Ad hoc On-Demand QoS routing* extends AODV in which a RREQ also includes the bandwidth and delay constraints. Intermediate nodes will only rebroadcast the packet if it meets the specified constraints. Bandwidth reservation is performed while the first packet is sent from the source using this path. A delay is also measured during the route discovery. So, if more than one route exists towards the destination, route with minimum delay is chosen.

### Multicasting

*Multicasting* is defined as the transmission of a message to a set of hosts in the network. The recipients of the message have a group identity. Multicasting plays a pivotal role in data exchanges and collaborative task execution in ad hoc networks, as it reduces the traffic overhead involved when the same data has to be sent to multiple destinations. However, multicasting presents several challenges in ad hoc network due to changing topology, nodes mobility, limited resources and hostile environment. The various schemes proposed in literature for multicasting can be classified as tree, mesh and hybrid multicast routing protocols (Junhai, Liu et al. 2008). In the *tree based approach*, a tree like data forwarding path is maintained which is rooted at the source node of the multicast session. Examples are adaptive demand driven multicast routing protocol and associativity based ad hoc multicast routing protocol etc. (Jetcheva and Johnson 2001; Royer and Toh 1999). A different approach is *group shared tree approach* where a single tree is used for a group instead of maintaining tree for each node. Each source first forwards the packet to the root of the tree which is then forwarded through the tree to multicast receivers. Examples of protocols in this category are STAMP and MZRP (Canourgues, Lephay et al. 2006; Zhang and Jacob 2004) etc. The problem with tree based protocols is that the mobility of the nodes along the tree may cause packet drops.

In a *mesh based approach*, a mesh like structure is maintained among the members. These protocols lead to higher connectivity among the nodes and thus perform reasonably well in mobility conditions. Examples of mesh based multicast protocols can be seen in (Inn and Winston 2006; Soon, Park et al. 2008). There are some *hybrid protocols* as well that combine the advantages of tree and mesh like structure (Biswas, Barai et al. 2004). Thus, these protocols are both robust against mobility and efficient in performing the multicast operation. Interested readers are referred to (Junhai, Liu et al. 2008; Badarneh and Kadoch 2009) for detailed surveys on multicast routing protocols.

### Addressing

As ad hoc network doesn't have any prior infrastructure, there has to be some mechanism to perform address assignments in a coordinated manner. The problem of addressing becomes complicated with network partitioning or merging, as the addressing algorithms should be smart enough to identify duplicate addresses in this situation. One of the simplest approaches to perform addressing is to maintain complete information about the internet protocol (IP) addresses allocated to other nodes called *state-full* approach (Nesargi and Prakash 2002). Any arriving node consults one of the reachable nodes called initiator for an IP address. The initiator assumes an address for the node and confirms it from other nodes by broadcasting that it is assigning this address to a new node. If the positive replies are received from all nodes, the allocation process is successful. In contrast, *state-less approaches* allow configuration of addresses locally (Chen, Fleury et al. 2009). Duplicate detection techniques are then used to resolve conflicts. *Hybrid schemes* are also possible, where a set of nodes tries to provide a conflict-free address according to its available knowledge. Duplicates undetected, are resolved afterwards (Li Longjiang 2009).

Some of the approaches for addressing work by *extending the classical dynamic host configuration protocol (DHCP)* to addressing. In this direction, Ancillotti, Bruno et al. (2009) proposed an addressing approach called Ad Hoc DHCP. It works by using the conventional DHCP scheme for hybrid ad hoc networks (ad hoc network with some infrastructure support). Any incoming node first discovers the nodes in its surroundings. It then selects a DHCP relay agent from its neighbors. The DHCP relay agent will forward the DCHP request of this node to known DHCP servers, thus a new address is assigned to the node.

An important aspect of addressing is to maintain the *scalability* of addressing process. Hussain, Saha et al. (2010) proposed a scalable approach to addressing where a set of servers are evenly placed in the form of grid topology. These servers ensure the identification of duplicate addresses in its vicinity, and thus guarantee balancing of addressing tasks to the servers.

### Transport protocol

A transport protocol provides end-to-end delivery, flow control and reliable transmission of data to other nodes in the network. There are two popular transport protocols used in wired networks. *Transmission Control Protocol (TCP)* is a connection oriented protocol that provides guaranteed delivery of data, while *Unified Datagram Protocol (UDP)* is proposed for real-time multimedia applications.

It has been observed that TCP protocol doesn't provide good performance in ad hoc networks. Several studies have been done on the performance of TCP over ad hoc networks. The problem arises due to the congestion control mechanism used by TCP. The TCP congestion control algorithm is based on additive increase multiplicative decrease strategy. So, whenever any packet's loss is observed (due to timeout or duplicate acknowledgment), TCP assumes it as a congestion and drastically reduces the window size. In wireless ad hoc network, the packet losses are often due to other reasons besides congestion i.e. attenuation, multipath fading, network partitioning and route failures etc. As the window size is tuned based on additive increase multiplicative decrease, it takes time for the window size to regain its original value. Hence, the overall throughput is severely affected.

Several TCP variants have thus been proposed for ad hoc networks. A survey on TCP variants have been provided in (Francis, Narasimhan et al. 2012). Among them include approaches that utilize network layer feedback to improve the congestion control algorithm. *TCP-F* works by distinguishing between a route failure and congestion (Chandran, Raghunathan et al. 98). By using cross-layer communication, whenever a link failure is detected, the routing agent sends an explicit congestion notification to the sender node. The sender then freezes its state and waits for a route reestablishment notification; the sender then resumes its proceedings. Similar schemes have been proposed in *ATCP* and *TCP Bus* etc. ATP utilizes the network layer information for startup rate estimation, congestion detection, and control and path failure notification (Sundaresan, Anantharaman et al. 2005). Changa, Kanb et al. (2012) proposed a cross-layer approach where physical and routing layers information is exploited by the TCP to discriminate between various types of network events like channel errors, buffer overflows and link layer contention etc. Besides network layer feedback, various proposals use the information available at transport layer to detect route failures (Hanbali, Altman et al. 2005). For example, *Fixed RTO* assumes a link failure when a timer expires two times consecutively, thus the RTO is not doubled. Similarly, *TCP-DOOR* identifies the link failure based on the out-of-order delivery events (Wang and Zhang 2002).

Unlike TCP, there has not been significant research done on UDP for ad hoc networks. Only a few UDP variants have been proposed for real-time data transmission over wireless and ad hoc network (Mao, Bushmitch et al. 2006; Yang and Zhang 2011). However, several studies have been done on analyzing the performance of UDP over ad hoc network. Most of these studies analyzed the behavior when TCP and UDP are competing (Rohner, Nordström et al. 2005). When TCP and UDP use a common link, UDP suffers from packet loss and interruptions are seen in TCP. Some experiments have also been done on optimum UDP packet sizes for ad hoc network (He, Ge et al. 2007).

### Security

Ad hoc networks are prone to a large number of security attacks due to environment where they are deployed as well as due to the absence of prior security infrastructure. A number of active and passive threats are possible i.e. include signal jamming, eaves dropping, impersonation and DoS attacks etc. (Wu, Chen et al. 2006). Since, these problems span across the whole protocol stack, security in ad hoc network is considered a multilayer issue. Let's discuss the various approaches that can be adopted to counter for security issues. A comprehensive discussion on security in ad hoc network has been provided in (Islam and Shaikh 2013).

To address the issues of signal jamming, techniques like frequency hopping spread spectrum (FHSS) and direct sequence spread spectrum (DSSS) can be employed (Wu, Chen et al. 2006). For resolving the security attacks at routing layer i.e. wormhole attack, black-hole attack and byzantine attack etc., different approaches have been proposed. Among them are techniques based on

*cryptography* to ensure the integrity of routing messages that are being exchanged. Examples of such protocols are SAODV, ARAN and Ariadne etc. (Islam and Shaikh 2013). Other approaches are based on *exploiting the routing header* for identifying any consistencies in the exchanged messages or *multipath routing* to recover a routing message tempered in transit by malicious nodes. In this direction, the approaches by Kurosawa, Nakayama et al. (2007), and Raj and Swadas (2009) utilized the sequence number present in the routing header to identify any consistencies and misbehavior in the network. In multipath routing protocol SPREAD, a message is decomposed in to multiple sub-packets which are retransmitted through various paths (Lou, Liu et al. 2009). At the receiving end, the packets are recombined. It is ensured that if a subset of the original message is tempered by intermediate malicious nodes, the original message can even then be reconstructed. To counter for various types of intrusion attacks and viruses etc., techniques based on firewalls, antivirus and intrusion detection have also been proposed in some researches (Wu, Chen et al. 2006).

## Cooperation

In ad hoc networks, the nodes are dependent on each other for their operations. For example, in multi-hopped routing, nodes are required to relay packets of other nodes. This interdependency requires some cooperation mechanism to ensure proper functioning of the network operations. Issues can arise when the nodes don't cooperate as the actions performed by a node for other users i.e. packet forwarding and resource sharing etc., call for consumption of bandwidth, data storage and power of the node. In addition, there are misbehaving nodes on the network that don't cooperate so that operation of the network can be jeopardized. It has been observed that the presence of the selfish nodes even in small amount, degrades the network performance significantly (Zayani and Zeghlache 2009).

There are two major approaches proposed in literature in this regard. The approaches belonging to the first category are based on *enforcing cooperation*. The nodes are monitored for their behavior and are punished in case of any misbehavior. Kwon, Lee et al. (2010) proposed a reputation based scheme where interaction among nodes is modeled as a Stackelberg game. The nodes reputations are determined based on their cooperation at each game. Nodes are encouraged to cooperate only with other cooperative nodes. Zayani and Zeghlache (2012) proposed a cooperation enforcement scheme based on the weakest link TV game principle. The nodes try to obtain a longest sequence of successful forwarding of packets. Misbehaving nodes are punished by reducing their utility value drastically.

The other category of schemes proposed for cooperation enforcement is *incentive based* mechanisms. In this direction, Eidenbenz, Resta et al. (2008) introduced a virtual currency scheme based. The intermediate nodes receive compensation in form of residual energy level. Zayani and Zeghlache (2009) proposed a fair and secure incentive based scheme where the nodes are rewarded or charged credits. A payment aggregation scheme is proposed to generate a single receipt for multiple packets thus reducing the network traffic. In Secure Incentive Protocol, a destination generates an acknowledgment to increment the credits of all the intermediate nodes (Zhang, Lou et al. 2007). This may cause fairness issues, when intermediate nodes don't get rewarded due to the loss of acknowledgments. The payer might have to pay more in case of loss of reward/receive packets or the data packet doesn't reach to the destination.

## Data management

In the last decade, the problem of data management has emerged as an important issue for ad hoc networks. As the devices are getting smaller and the peer to peer connectivity mediums becoming smaller, the amount of data exchanges in the network is becoming huge. It is therefore imperative to have novel data management mechanisms that can analyze the gigantic volume of data to extract meaningful information. Such a system should address the aspects related to acquisition, representation, storage and protection of data etc. The problem arises in data management because of the limited storage, lack of global schema, heterogeneity and spatiotemporal variation in states of data etc. Various data management systems have been proposed in recent years. Among them are MoGATU (Perich, Joshi et al. 2006), DRIVE (Zhong, Xu et al. 2008), CDMAN (Martin and Demeure 2008) and CHaMeLeoN (Shahram Ghandeharizadeh 2006) etc. A survey of data management frameworks is provided in (Islam and Shaikh 2011).

### Testbeds and simulation

As newer algorithms and approaches are being proposed for ad hoc networks, a deep analysis of them is required before their deployment. Various approaches have been adopted in this regard. One approach is *analytical modeling*, which is not viable because the complex nature of ad hoc network can't be modeled precisely. The second approach is to use *testbeds*. Several testbeds are available for ad hoc networks (Kiess and Mauve 2007). However, testbed also suffers from different economical, monitoring and implementation issues. The third approach is the *simulation* of the algorithm in a software environment. There has been a number of commercial and open source software proposed for ad hoc network. A summary is provided in Table 1.

*Table 1: A summary of various ad hoc network simulators*

| Simulator | Free | Open source | Language | Features | Website |
|---|---|---|---|---|---|
| NS-2 | √ | √ | C, TCL | TCP, Routing protocols, Multicasting, Visualization support | http://www.isi.edu/nsnam/ns |
| OMNET++ | √ | √ | C++ | Wireless and mobile network simulation | http://www.omnetpp.org |
| Glomosim | √ | ✕ | Parsec | Scalable Simulation, support for various types of mobility models, visualization, ad hoc networking | http://pcl.cs.ucla.edu/projects/glomosim |
| SWANS++ | √ | √ | Java | Support for wireless and ad hoc networks, routing protocols, mobility models | http://sourceforge.net/projects/straw |
| Opnet | ✕ | ✕ | C++ | VoIP, TCP, OSPF, ad hoc networks, IPv6, grid computing | http://www.opnet.com |

### Mobility models

*Mobility models* are used to describe the movement of real-world entities in to a simulation environment. Mobility models proposed for ad hoc networks can be categorized as *individual mobility* and *group-based* mobility models (Camp, Boleng et al. 2002). In the former case, the mobility patterns of a node are independent of other nodes; while in the later case, the mobility of a particular node is dependent on other nodes' mobility. Following are some of the most popular mobility models proposed in literature (Cooper and Meghanathan 2010):

- **Random Walk:** A node randomly chooses a direction and speed, continues moving until a particular destination is reached or a certain time has elapsed
- **Random Way Point:** A node selects a target and speed, moves towards the target, pauses for some time and then reselects a new target
- **Random Direction Walk:** A node walks randomly in a particular direction with random speed, until it reaches the boundary of simulation area, where it pauses for some time and then reselects a new direction
- **Gauss Markov Move:** The node speed and direction is a function of the last move and direction
- **Probabilistic Random Walk:** Three states are defined for movement in x and y axes. In x-axis, the node can move left, right or still. Similarly in y-axis, up, down or still states are possible. The states of the node can also be changed based on some probability

- **Column mobility:** All the nodes move randomly, except a subset of nodes. Among this subset, there is a leader and rest are followers. The followers move behind the leader in a lane
- **City section:** The node movement is constrained to a grid like structure comprising different streets. There is a speed limit for each street and the nodes' movements in a street are within these speed limits. Nodes are place on a particular intersection at the start of simulation. Node then moves to a random intersection with at most one move. The process is repeated until the simulation ends
- **Manhattan model:** Similar to city model, except a probabilistic model is employed to select the subsequent street in which the node can move

Besides these simple mobility models, another approach is to use specialized tools to generate mobility patterns. Various tools have been proposed for this purpose. For example, MOVE is a Java based toolkit that allows the user to quickly generate mobility patterns by using its map editor or importing the maps from other databases (Karnadi, Mo et al. 2007).

*Table 2: A summary of research issues in ad hoc network*

| Research Issue | Issues/ Challenges | Solutions | Reference |
|---|---|---|---|
| Antenna Design | Channel capacity, multipath effect, interference | Smart Antenna | (Chau, Gibbens et al. 2012) |
| Energy management | Low powered nodes, coordinated energy management at various layers | Transmission power control Energy efficient routing | (Singh and Kumar 2010) (Liu, Guo et al. 2009) |
| MAC Layer | Hidden node, exposed node problems, deafness, head of line blocking | Contention free and contention based solutions QoS aware MAC MAC for directional antenna | (Kumar, Raghavan et al. 2006) (Wang, Zhai et al. 2008) (Kamruzzaman, Hamdi et al. 2010) |
| Routing | Blackhole attack, worm hole attack, routing loops | Proactive, Reactive, Hybrid routing protocols Position based routing | (Boukerche, Turgut et al. 2011) |
| Multicasting | Changing topology of the network, resource limitations | Tree, mesh and hybrid protocols | (Junhai, Liu et al. 2008) |
| Addressing | No addressing server, possible DoS attack | State-less and State-full approaches | (Nesargi and Prakash 2002) (Chen, Fleury et al. 2009) (Li Longjiang 2009) |
| Transport Protocol | Low throughput due to congestion control | Network layer information utilization | (Francis, Narasimhan et al. 2012) |
| Security | Hostile environment, lack of security infrastructure | Secure routing, intrusion detection | (Lou, Liu et al. 2009) (Svecs, Sarkar et al. 2010). |
| Cooperation | Lack of trust among nodes | Cooperation enforcement Incentive based mechanism | (Kwon, Lee et al. 2010) (Zayani and Zeghlache 2012) |
| Data Management | Data interoperability, data discovery | Semantic representation, cross-layer information exchange, opportunistic data sharing | (Martin and Demeure 2008) (Islam and Shaikh 2011) |
| Testbeds and simulation | Network complexity, cost of hardware | Test beds Simulator | (Kiess and Mauve 2007) |
| Mobility models | Realistic simulation of mobile nodes | Mobility models Patterns generator | (Karnadi, Mo et al. 2007) |
| Standardization | Lack of harmony among standardization bodies | IEEE , IETF | (IETF 2012) |

### Standardization

Although lot of research efforts are underway for various issues of ad hoc network as discussed above, there is a need for standardization for its wide range acceptance. Even though there have been various standardization bodies working on the topic. For example, IETF MANET working group is working on standardization of routing protocols (IETF 2012). Various IEEE wireless standards are available for use in ad hoc networks (Jangra, Goel et al. 2010). However, all of these standards are focused on a particular aspect. There is a need for standardization that looks at the global picture. In particular, efforts are required on evolution of architecture, open standards and protocols. These efforts should also have provision for incorporating current efforts and also integrate current wireless and cellular networks.

Table 2 provides a summary of different research issues of ad hoc network. It can be seen that different type of challenges arises when addressing various issues. The different solutions proposed to address challenges are summarized in the table. However, current solutions are still immature and a lot of research areas are still open, as we will see in later sections.

## EMERGING MODELS OF AD HOC NETWORKS

During last few years, different emerging modes of ad hoc networks have been proposed. These variants emerged as applications of ad hoc networks in specialized domains or due to their intersection with other technologies. Figure 1 provides five broad categories of ad hoc networks based on their characteristics. A *sensor network* comprises a number of miniscule sensing devices that organize themselves autonomously for monitoring physical environments. These networks are useful for environmental monitoring, body area networks, smart office, smart agriculture and precision farming etc. A *mobile ad hoc network (MANET)* is a type of ad hoc network with mobile nodes. As the nodes are mobile and the network is highly dynamic, links failure and topological changes are common phenomena in such networks. When the ad hoc network is created among vehicles, it is called *vehicular ad hoc network (VANET)*. These networks are attributed by highly mobile nodes, more processing capabilities and availability of location tracking devices etc. VANET has been used for the management of various transportation problems (Islam, Shaikh et al. 2008). Another type of ad hoc network is *wireless mesh network (WMN)* that is formed among a set of self organizing nodes arranged in a mesh topology. WMN provides more resilience and cost-effective connectivity and has been used for military operations, community networking and telemedicine etc. (Akyildiz, Wang et al. 2005). In the last few years, techniques have been proposed to optimally utilize unused licensed spectrum available around a node. *Cognitive Radio Ad hoc Network (CRAHN)* comprises various self-organized nodes that are equipped with cognitive radios that utilize unused portion of spectrum for internal communication.

*Grid computing* is the concept of aggregating a collection of distributed loosely couple resources. By integrating with ad hoc networks, the concept of *ad hoc grid* has emerged. Similarly, the integration of sensor network and grid computing gives rise to concept of *sensor grid*, where the data collected from sensor networks are passed to a grid for high performance computing. In some of the researches, *vehicular grid* has been proposed as an integration of vehicular network and grid computing. *Cloud computing* is also an emerging discipline where the resources are leased from a collection of tightly coupled and virtualized resources. The merger of ad hoc networks and cloud computing has given rise to the notion of a*d hoc clouds*. Besides, the concept of mesh clouds has also been proposed in some researches.

In addition to the various models discussed above, other novel network models are also possible. Examples are sensor clouds, cognitive vehicular networks, cognitive clouds, cognitive sensor networks, cognitive vehicular network and cognitive wireless mesh networks etc. (Akan, Karli et al. 2009; Felice, Doost-Mohammady et al. 2012). In the sections below, the discussion has been restricted to ad hoc grids, ad hoc clouds, wireless mesh networks and cognitive radio ad hoc network. We will discuss briefly these models, their attributes and additional research issues arise during their implementation. The various solutions to address these issues are presented very briefly due to space limitations. However, pointers to corresponding literature are provided to interested readers for further research.

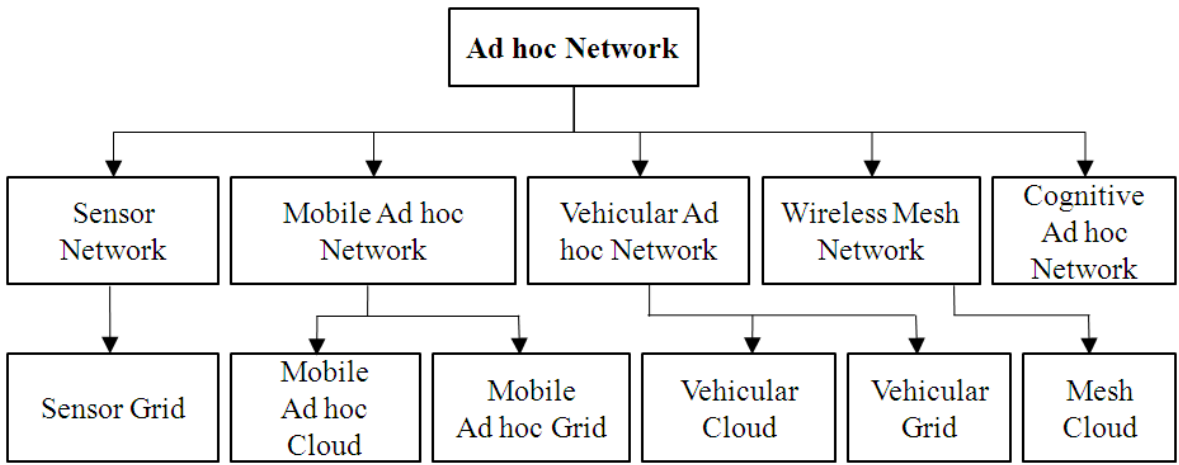

*Figure 1: A classification of various types of Ad hoc Networks*

## Ad hoc grid

*Grid Computing* is an approach towards accumulation of a number of distributed and loosely coupled resources for coordinated problem solving. The central idea of grid computing is to provide the resources i.e. hardware, software, data and network etc. as a utility similar to conventional electricity grid. By the sharing and optimal utilization of distributed resources, grid computing has enabled large scale computing that was previously possible only with super computers. With the enormity of computing devices around, the potential of integrating grid computing and ad hoc networks has been studied in recent years. This leads to the concept of *ad hoc grid* where various mobile devices organize themselves on the fly in the form of grid.

Due to recent advances, the commonly used appliances in our daily lives i.e. microwave ovens, wrist watches, glasses and air conditioner etc. are expected to behave like computing gadgets. Generally, these resources remain unutilized most of the time. The concept of ad hoc grid is to optimally utilize the capabilities of these devices by self organizing them for different computationally intensive tasks. Ad hoc grids possess all the properties of ad hoc network i.e. self organization, autonomy, distributed operations, multi-hop routing, and mobility of nodes etc. However, they also have some additional properties that set them apart for conventional ad hoc networking. Let's discuss them briefly.

### Properties of ad hoc grid

As the grid computing enables resource sharing across organizational boundaries. The different resources available are *heterogeneous* in terms of architecture, capabilities, performance and operating environment etc. The resources can belong to *different administrative domains* or even *geographically distributed*. Ad hoc grids are generally used for solving *large scale* problem. Since, the overall objective of grid computing is to provide resources; *fault tolerance* is an important requirement for any ad hoc grid system. The nodes are required to have better *coordination* and therefore mechanisms for cooperation enforcement is a more stringent requirement. Finally, the consumers of services should provide *transparent access* above the heterogeneous and dynamically changing ad hoc environment.

### Research issues in ad hoc grid

Ad hoc grid presents newer challenges in addition to the one posed by ad hoc network. New *resource discovery algorithms* are required for ad hoc grids that should provide the ease of deployment of ad hoc network coupled with fault tolerance and scalability demands of grid environment. A resource discovery scheme should be energy efficient and takes less time to discover resources. Conventionally, the resource discovery algorithms have been classified as

*directory-based* and *directory-less approaches* (Islam, Shaikh et al. 2010). The former utilizes a directory for maintain list of available resources in the network. For instance, Mallah and Quintero (2009) proposed a directory based approach in which a set of backbone nodes are selected for maintaining directories and entertaining discovery requests. Directory-less approach works in decentralized fashion by contacting the hosting nodes for the desired resources. In this direction, Pariselvam and Parvathi (2012) proposed a directory less scheme based on swarm intelligence to perform resource discovery in ad hoc networks. A similar approach has been proposed by Singh and Chakrabarti (2013), where super nodes are selected based on ant colony optimization. These super nodes are responsible for discovery of resources in the grid. Various cross layer discovery scheme have been proposed in recent literature that merges the discovery process with other protocols. For example, Islam and Shaikh (2013) presented a cross-layer approach to service discovery based on integration of discovery process with routing protocols. A secure cross-layer service discovery SCAODV has also been proposed by (Zhon, Geng et al. 2012). Cross-layer solution still remains an unexplored area in the domain of ad hoc grids, however.

Another important research issue in ad hoc grid is the development of *scheduling algorithms* to allocate certain type of resources for the execution of a particular job. For ad hoc grid, scheduling algorithms are required that should not only consider the classical parameters but also consider the attributes reflecting the dynamism in the network. This includes the nodes mobility, failures of links and nodes, and network partitioning etc. There are three types of scheduling schemes proposed in literature. The *centralized approach* is based on a focal entity responsible for scheduling the resources. The *decentralized approach* comprises different schedulers that interact with each other for scheduling of resources. Finally, the *hierarchical scheduling* is based on several low level schedulers and a top-level scheduler for global coordination. Accroding to Bhaskaran and Madheswaran (2010), a decentralized job scheduling is better suited for addressing the fault tolerance and reliability of ad hoc grids. They discussed various challenges in implementing scheduling in ad hoc grids. The authors also proposed a scheme that works in decentralized fashion to perform job scheduling and congestion control in ad hoc network. Among other works, Torkestania (2013) presented a distributed job scheduling algorithm in which the jobs are allocated to a node based on its computational capacity. The schedulers are synchronized by a learning automata. Xu and Yin (2013) have also proposed a task scheduling algorithm based on a mathematical model. The scheme considers the mobility of nodes and resources while performing task scheduling.

Relevant to scheduling, is the development of robust *work flow systems* that can compose complex tasks and schedule its execution in the ad hoc environment (Terracina, S. Beco et al. 2006). The currently available workflow systems can be classified as *non context-aware* and *context-aware* workflows. Among the non-context aware workflow systems is Grid Services Flow Language (GSFL), proposed by Krishnan, Wagstrom et al. (2002). GSFL is essentially a non-context-aware workflow description language to represent service providers, activity model, composition model and lifecycle model for implementation of workflows in a grid. However, Abbasi (2013) hypothesized that a context-aware workflow system has more capabilities to adapt for ad hoc changes in the environment. He presented a context-aware approach to design and manage workflows, comprising specialized context-aware activities that can be dynamically loaded based on the context. A context-aware work flow management system for ubiquitous applications has also been presented by Tang, Guo et al. (2008).

*Security and trust management* are essential components of an ad hoc grid system. A security system should provide authorization, protection and establishing the trust among peers to ensure collaboration in a fair manner (Amin, Laszewski et al. 2004). Different security solutions have been proposed for ad hoc grid systems based on trust management. For instance, RETENTION is a reactive trust management scheme based on a mathematical trust model in which malicious nodes are punished for their misbehavior (Bragaa, Chavesb et al. 2013). Singh and Chakrabarti (2013) presented a trust management scheme for mobile grid system. There are super nodes in the network elected based on different parameters i.e. CPU, battery power and bandwidth. For ensuring the trust among nodes, a trust collection mechanism has been adopted. Huraj and Siládi (2009) have presented a survey of various authorization and authentication shames for ad hoc grids and also presented a security scheme based on trust chains developed using public key infrastructure.

Another important issue in ad hoc gird is the design of *economic models* such that services can be fairly utilized in the network. There should be some utility value associated with a service and a billing mechanism should provide incentives for services provided by a node. A similar work has been cited in (Vazhkudai and Laszewski 2001), where a basic framework is proposed comprising services for bartering, bidding and trading etc. Li and Li (2012) also presented an ad hoc grid system based on micro economic theory where the producers and consumers are modeled as decision makers who buy and sell resources similar to economic architecture.

New *QoS models and metrics* are required for ensuing QoS in ad hoc grids. In general, the current QoS approaches for grid can't cater with dynamism and the new requirements for collaboration and resource sharing posed by ad hoc grids (Li, Sun et al. 2005). Among the few solutions proposed, Li and Li (2012) have presented a scheme for QoS and load balancing. A utility values is associated with the QoS satisfaction and a preference value is computed from the resource point of view. These two values are combined to select a resource for a particular request.

Unlike conventional grid computing system, an ad hoc grid comprises mobile nodes that have limited battery life. Hence, the architecture and solutions proposed for ad hoc grid should be *power efficient*. In this direction, Marinescuand, Marinescuand et al. (2003) proposed a basic architecture for ad hoc grid and then presented a model for power consumption for different operations of the network. Shah, Bashir et al. (2009) have also presented an architecture for establishment of ad hoc grid among resource constrained devices. The architecture takes care of energy limitations, mobility and signal strength issues. There is a configuration profile service running that keeps track of battery power and report to grid job scheduler to switch a job to some other node.

Table 3 summarizes the various research issues related to ad hoc grid as discussed in previous paragraphs. In some of the literature, different sub-types of ad hoc grids systems have also been proposed. This section is concluded with brief discussions on these models.

*Table 3: Summary of research issues in ad hoc grid*

| | Research issue | Challenges / Issues | Representative Approaches |
|---|---|---|---|
| 1. | Resource discovery | Limited battery | Directory-based |
| | | Lack of global schema | Directory-less |
| | | Heterogeneity | Cross-layer (Islam and Shaikh 2013) |
| 2. | Scheduling | New scheduling criteria | Centralized scheduling |
| | | Node mobility | Hierarchical scheduling |
| | | Failures | Distributed scheduling  (Bhaskaran and Madheswaran 2010; Torkestania |
| | | Network partitioning | 2013; Xu and Yin 2013) |
| 3. | Workflow | Network dynamism | Context-aware systems (Abbasi 2013; Tang, Guo et al. 2008) |
| | | Context management | Non context-aware system (Krishnan, Wagstrom et al. 2002) |
| 4. | Security | Authentication and | (Bragaa, Chavesb et al. 2013) |
| | | authorization | (Singh and Chakrabarti 2013) |
| | | Trust establishment | (Huraj and Siládi 2009) |
| 5. | Economic model | Utility assignment | (Vazhkudai and Laszewski 2001) |
| | | Faire utilization of resources | (Li and Li 2012) |
| | | Ensuring cooperation | |
| 6. | Power control | Mobility | (Marinescuand, Marinescuand et al. 2003) |
| | | Signal variations | (Shah, Bashir et al. 2009) |
| 7. | Quality of service | New metrics and models | (Li and Li 2012) |
| | | considering collaboration and | (Marinescuand, Marinescuand et al. 2003) |
| | | resource sharing | (Shah, Bashir et al. 2009) |

### Sensor grid

*Sensor grid* emerged as an integration of sensor network and grid computing system. In a sensor grid, the data collected from sensors is sent to a grid environment for data processing and actuation. Sensor grids have been applied for solving different types of applications. For example, *AgriGrid* is a context-aware sensor grid platform proposed for management of agriculture problems (Aqeel-ur-Rehman and Shaikh 2008). Similarly, *iHEM* is a sensor grid based approach for in-home management of electrical energy (Erol-Kantarci and Mouftah 2011). It is based on wireless sensor based communication among the consumers of electricity and the controllers.

Sensor grid presents several new research issues besides the challenges associated with sensor network and grid computing environment. An overview of various issues is provided in (Tham and Buyya 2005). One of the core issues is design of *service oriented approaches* for access, distributed processing and management of sensors. SensorML is an effort in this direction that provides XML schema for defining sensor characteristics (OpenGeospatialConsortium 2013). Efficient and robust *querying* mechanism are also required for sensor grid to fetch data in real-time from the distributed sites by sensors and vice-versa. Novel protocols are required to deal with *security*, *network failures* and *noise issues* that arise due to wireless nature of the network. Finally, mechanisms are required for *addressing*, *scheduling*, *power management* and *QoS provisioning* in sensor networks. Representative solutions can be seen in (Miridakis, Giotsas et al. 2009; Li, Li et al. 2013).

### Vehicular grid

*Vehicular grid* is a grid environment established among a number of vehicles travelling on the road that are connected in ad hoc fashion. Different vehicular grid frameworks have been reported in literature. Khorashadi (2009) proposed a vehicular grid framework called *vGrid* in which the vehicles on the road share their information with other vehicles to solve various traffic management problems. The current ITS solutions for traffic management are very expensive and require large amount of time for solving traffic problems. According to Khorashadi (2009), by employing the vehicular grid solution, the traffic problems can be solved in order of seconds. The various types of problems that can be solved using the proposed framework are lane merging, accidental warning and ramp metering etc. In the same direction, Islam, Shaikh et al. (2008) proposed a grid based approach to routing traffic in vehicular ad hoc networks. By establishing a grid between vehicles, a distributed shortest path algorithm is used to find out optimal routes. Another relevant project is *VGITS*, a hybrid system based on data processing and real-time traffic services in a centralized manner, while providing traffic services to drivers in a decentralized manner (Chen, Jiang et al. 2008).

Beside the various issues of ad hoc grid, vehicular grid systems present several additional challenges. These are mainly related to development of a standards, architecture, security and privacy, routing protocols, proposal for novel applications, and development of testbeds and toolkits etc. for implementation of algorithms and protocols.

## Ad hoc cloud

*Cloud computing* is a distributed computing paradigm in which a pool of computing, storage and platforms are delivered as a service over the internet. Grid and cloud computing have several things in common i.e. utility computing, service oriented architecture and distributed operation etc. However, grid comprises loosely coupled resources spanning across multiple organizational boundaries and cloud is a collection of tightly coupled resources virtualized and available for user as service over internet. The examples of various cloud computing environments are *Amazon Elastic Cloud Computing*, *Microsoft Windows Azure* and *Google App Engine* (Islam and Aqeel-ur-Rehman 2013).

The motivation behind cloud computing is to address resource limitation problems of mobile computing devices. The integration of cloud and mobile computing has been studied recently to address the resource limitations in ad hoc environments. Kovachev, Cao et al. (2011) have provided a survey of various models for *mobile cloud computing*. One of the approaches is offloading the task of a mobile device to a cloud. Another approach is *ad hoc cloud* where a set of mobile devices forms a cloud computing environment and provides services to other devices. Ad hoc cloud can be used for different types of applications ranging from cloud robotics, crowd computing and data sharing to image processing etc. (Hu, Tay et al. 2012; Fernando, Loke et al. 2013).

## Properties of ad hoc clouds

The resources in ad hoc clouds are provided as different types of *services* to end users. These services models can be infrastructure, platform, software, network or storage as service (Islam and Aqeel-ur-Rehman 2013). Unlike ad hoc networks, the resources are *pooled* and provided to consumers as *virtualized* resources. Ad hoc clouds offer a more flexible *on-demand servicing* approach where the resources are *dynamically provisioned* with the increase in demands of users. Contrary to conventional clouds, the nodes in ad hoc cloud are not dedicated, instead every node share their computational resources. Hence, there must be some mechanism for *measurement* of services and users should be provided appropriate incentives for the resources shared. Ad hoc clouds have relatively *more resources* and are therefore useful for computationally intensive tasks. Finally, ad hoc cloud should be *fault tolerant* and provide *transparent access* of services to end users irrespective of failure of certain services in cloud.

## Research issues in ad hoc clouds

There are various research issues associated with implementation of ad hoc cloud as discussed in (Kirby, Dearle et al. 2010; Fernando, Loke et al. 2013). One of the issues is to decide the mechanism to *offload a task* for execution on cloud. Hu, Tay et al. (2012) have discussed various issues to be considered while offloading a task for execution. Among the strategies are remote procedure call (RPC) or remote method invocation (RMI) to upload the task, and migrating the virtual machine (Fernando, Loke et al. 2013). There are a variety of issues related to *service provisioning* in ad hoc cloud. For example what are the service models supported by the cloud and how the cloud ensures smooth provisioning of these services in failure? The first question is dependent on the application domain. In this direction, Olariu and Yan (2012) have presented a vehicular cloud system in which the services supported are related to traffic management problems. In case of disconnection from mobile cloud, a vicinity node connected to the cloud can be consulted for the service (Dinh, Lee et al. 2011). Another relevant issue is the provision of QoS of the services (Zhang and Yan 2011). To ensure elasticity of services offered on cloud, an application model has been proposed in (Zhang, Schiffman et al. 2009).

Buyya, Yeo et al. (2009), on his vision on 21$^{st}$ century of computing stressed on the need for marked oriented resource management for cloud computing and presented an architecture for supporting market oriented resource allocation. Hence, an important issue in ad hoc clouds is the availability of an *economic model*. In this direction, Liang, Huang et al. (2012) presented a model based on Semi-Markov decision process. The model considers the maximal system rewards and expense of mobile device to devise an optimal resource allocation policy. *Energy efficiency* is becoming a fundamental issue in cloud computing. Novel mechanisms are required to ensure minimal energy consumption during their operation. Miettinen and Nurminen (2010) presented a model for ad hoc cloud and highlights critical factors responsible for energy consumption in clients in mobile clouds. Some researchers have considered network coding and cooperation among the nodes as mechanisms for energy conservation in mobile clouds (Heide, Fitzek et al. 2012).

Another issue is related to the *interoperability* and heterogeneity of devices that are in used on the cloud (Sanaei, Abolfazli et al. 2013). Unfortunately, there are no open standard leading to scalability, deployment and availability of service issues (Chetan, Kumar et al.). The current web standards are not useful as they were not designed for mobile devices. HTML5 offers web-sockets that provides real-time communication and can serve as a useful candidate for ensuring interoperability (Dinh, Lee et al. 2011).

There are a number of issues pertaining to *security* in ad hoc clouds. This includes integrity and privacy of data shared on the cloud, authentication of clients that can use these services and establishment of trust among the nodes in the network. Popa, Cremene et al. (2013) proposed a framework for mobile clouds based on applying security policy depending on the type of data. The security is ensured during communication, and the integrity of the applications are also ensured during their installation and updating process. Discussions on various public key cryptography schemes for mobile cloud have been provided in (Zheng 2013). In addition, a general comparison of various security schemes for ad hoc clouds has been reported in (Iyer and Durga 2013).

Table 4 provides a summary of major research issues in ad hoc cloud and representative approaches taken to address them. Besides these approaches, various ad hoc cloud frameworks been proposed in literature to address the various issues. Huerta-Canepa and Lee (2012) presented a preliminary design for creation of ad hoc cloud computing solution. Remote procedure calls are used for offloading of task based on current context. A resource management component performs profiling and monitoring of devices. Harox is another framework that provides an approach towards cloud services on android phones in which various benchmark algorithms of distributed sorting and searching are implemented using Hadoop framework (White 2012; Marinelli 2009).

*Table 4: Summary of research issues in ad hoc cloud*

| | Research issue | Challenges / Issues | Representative approaches |
|---|---|---|---|
| 1. | Offloading of task | Heterogeneity<br>Distance | Remote procedure call<br>Remote method invocation<br>Virtual machine migration |
| 2. | Service provisioning | Service models<br>Link failure<br>QoS<br>Dynamic provisioning | Network, storage, coordination as services (Olariu and Yan 2012)<br>(Dinh, Lee et al. 2011)<br>Zhang and Yan (2011)<br>(Zhang, Schiffman et al. 2009) |
| 3. | Economic model | Utility assignment<br>Negotiation<br>Payment mechanisms | (Buyya, Yeo et al. 2009)<br>(Liang, Huang et al. 2012) |
| 4. | Energy management | Limited resources<br>Energy aware algorithms | Network coding, node cooperation (Heide, Fitzek et al. 2012) |
| 5. | Interoperability | Heterogeneity<br>Lack of standardization | Websockets (Dinh, Lee et al. 2011) |
| 6. | Security | Integrity<br>Privacy<br>Authentication<br>Trust management | (Popa, Cremene et al. 2013)<br>Cryptography (Zheng 2013)<br>(Iyer and Durga 2013) |

### Vehicular cloud

A sub-discipline of ad hoc cloud is *Vehicular Cloud Computing*, proposed recently in (Olariu and Yan 2012). The motivation was that vehicles on the street, fuel stations, and parking places of offices, restaurant and shopping plazas etc. have huge underutilized computational resources. These resources include on-board computer, storage, location tracker, communication devices and camera etc. If these resources can be organized in the form of a cloud, they can be optimally utilized for solving various types of traffic problems. The author proposed different types of cloud service models. For example, internet connectivity available at some nodes can be rented out to other nodes to give rise to concept of *network as a service*. Similarly, nodes can share their extra storage for other nodes that needs large storage for their applications i.e. *storage as a service*. Finally, nodes can cooperate with each other to provide various ITS services. This leads to the concept of *cooperation as a service*.

Let's discuss representative proposals appeared in literature on vehicular cloud. Wang, Cho et al. (2011) proposed a vehicular cloud system for improving driving comfort and providing different types of services to cloud. A layered architecture is presented, various types of applications and services are discussed, and research challenges are also highlighted. Hussain, Son et al. (2012) have provided taxonomy of various vehicular cloud models and grouped them in to vehicular clouds, vehicles using clouds and hybrid vehicular clouds. Alazawi, Altowaijri et al. (2011) have presented a slightly different approach for disaster management system. It is based on gathering information from multiple sites, which is then processed on a cloud. The point of incident is identified and effective strategies to counter them are devised. This information is then disseminated to vehicles.

Vehicular clouds can play a key role in disaster management, intelligent transportation and similar applications, especially in developing countries. However, several new research challenges needs to be addressed. Among them, some of the challenges are related to information dissemination, standardization, privacy, authentication, trust management and availability of hardware in vehicles etc. (Talebifard and Leung 2013; Olariu and Yan 2012; Yan, Rawat et al. 2012).

## Wireless mesh networks

*Wireless mesh network (WMN)* is a self organized network established among a set of stable nodes arranged in mesh topology to provide network access and other types of services to mobile clients. The former set of nodes is called *mesh router* that serves the later type of nodes called *mesh clients*. WMN are generally useful for interconnecting different wireless networks with quick deployments and low upfront costs, and gives rise to several useful applications as discussed earlier.

### Properties of wireless mesh network

Sichitiu and Akyildiz, Wang et al. (2005) have discussed various key properties of WMN. One of the important characteristics of WMN is *integration* of different wireless networks. The nodes in WMN are of three different types. There are *mesh routers* that are less mobile and provides services to *mesh clients*. Mesh clients can be stationary or mobile. Some of the mesh nodes serve as *internet gateways* providing interconnectivity among different networks. This also raises the need for countering the *heterogeneity* among these networks. WMN operates in multi-hopped fashion providing interconnection in short distances. Additionally, to increase the effective bandwidth, some nodes have *multiple radio interfaces*, allowing them to work on different channels.

Since, WMN comprises a wireless backbone infrastructure with mesh routers, there are *no power management* issues. The network is *scalable* and the load on mesh client side is very minimum. This allows limited capability devices to be part of the network. Most of the operations i.e. routing and configuration etc. are performed by routers and there is a *steady communication* among the mesh routers. Finally, WMN can also be deployed in *preplanned or incremental* fashion in some situations.

### Research issues in WMN

Because of the characteristics of WMN, different new challenges and issues surfaced out at various layers of protocol stack (Akyildiz and Wang 2005; Pathak and Dutta 2011). An important research issue is *modeling* of the various parameters of WMN i.e. capacity, data rate and flow assignment etc. (Rawat, Zhao et al. 2013). The availability of an analytical model provides basis for further analysis and promotes the development of new architecture, applications and protocols for WMN. In this direction, Dely, Castro et al. (2010) analyzed the throughput of a multi-radio mesh networks under different conditions. The authors concluded that for low PHY rate, channel separation provides hints for channel capacity. On the contrary, under high PHY rate, channel capacity measurement requires analysis of propagation properties. Research efforts are also required to develop *testbeds and simulators* to validate research proposals. Representative works have also been reported in (Rawat, Zhao et al. 2013; Lam, Busan et al. 2012).

At physical layer, advanced techniques are required to enhance the *transmission rate* of the channel. This includes novel antenna designs techniques i.e. diversity coding, smart antenna and directional antenna etc. The various techniques that can be employed for improving the reliability and data rate of transmission are orthogonal frequency multiple access, code division multiple access and ultra wide band techniques (Sichitiu 2005).

As the WMN comprises mesh client and routers, new *MAC protocols* are required that can work both for mesh clients and routers. MAC protocols should also be able to work under different wireless technologies and capitalize on multiple communication channels. A simple approach is to use a dedicated channel for controlling. However, the control channel can get saturated leading to non-optimal channel utilization. To address this problem, Lei, Gao et al. (2012) proposed a wait-time based approach to multi-channel MAC protocol. When directional antennas are used in WMN, new MAC protocols are required that should counter different problems as discussed in previous section. A survey of MAC protocols for directional antenna is provided in (Bazan and Jaseemuddin 2012; Lu, Towsley et al. 2012). *Scalability* is also an important requirement for multiple access protocols for WMN. Unfortunately, most of the MAC and routing protocols for ad hoc network don't cater with the scalability requirements and only a few scalable MAC protocols are currently available (Zhou and Mitchell 2009). Therefore, designing scalable MAC protocols for WMN is an important area of research.

The *routing protocols* for WMN pose stringent requirements as compared to ad hoc networks. This includes enhanced scalability, fault tolerance, load balancing, power control, and adaptability to work for mobile clients, (Sichitiu 2005). Unlike the conventional metrics, new metrics should be used to meet these requirements. A routing protocol should combine more than one metric considering the transmission count, round trip time, noise and interference to compute routes. An overview of different routing protocols and metrics for WMN has been provided in (Campista, Esposito et al. 2008). The authors classified current protocols as those based on reduced link variations, control overhead and network traffic, and opportunistic routing. To enhance the reliability, multi-path routing scheme has also been proposed in some research (Akyildiz, Wang et al. 2005).

The design of *transport layer protocols* also presents several new challenges in mesh networks (Rangwala, Jindal et al. 2008). It should have the flexibility to operate under different network and protocols running at lower layers. Depending on the network, there can be different flow control and error control parameters. Therefore, transport protocols should be able to cater to these variations. As discussed in previous section, TCP suffers from various issues and a range of TCP variants have been proposed. Similarly, variants of UDP i.e. RCP/ RTCP have also surfaced to provide reliable real-time communication. A survey of various transport protocols for wireless mesh networks have been provided in (Law 2009).

Like ad hoc networks, WMN also suffers from various security issues. This includes eaves dropping of mesh management and control frames, jamming communication among mesh routers and masquerading as mesh nodes by an intruder etc. (Zdarsky, Robitzsch et al. 2010). Among the various approaches proposed to security are mobile client authentication and access control techniques, cryptographic approaches to secure communication, securing mesh routing and securing backbone etc. (Zhang and Y.Fang 2006; Martignon, Paris et al. 2011). The *application layer* of WMN also requires careful design of applications such that they can adapt themselves under the architecture of WMN.

Table 5 provides a summary of discussed research issues in WMN. Besides discussed, there are a number of other research issues that require investigation i.e. cross-layer design, multicasting, network planning and deployment etc. (Akyildiz, Wang et al. 2005).

*Table 5: Summary of research issues in wireless mesh networks*

| | Research issue | Challenges / Issues | Representative approaches |
|---|---|---|---|
| 1. | Implementation | Channel modeling<br>Testbeds and simulators | (Dely, Castro et al. 2010)<br>(Rawat, Zhao et al. 2013; Lam, Busan et al. 2012) |
| 2. | Transmission rate | New approaches to increase capacity<br>Address noise and interference | New modulation and channel coding techniques<br>Multiple array antenna techniques |
| 3. | MAC protocols | Heterogeneity<br>Multi-channel MAC<br>Scalability | (Lei, Gao et al. 2012)<br>(Zhou and Mitchell 2009). |
| 4. | Routing protocols | New metrics considering fault tolerance and power efficiency | Ad hoc based<br>Controlled flooding<br>Traffic aware<br>Opportunistic routing<br>Multipath routing |
| 5. | Transport protocols | Increased packet loss<br>Heterogeneity in error/ flow control | (Law 2009) |
| 6. | Security | Mobile client authentication<br>Access control<br>Secure routing | (Zhang and Y.Fang 2006)<br>(Martignon, Paris et al. 2011) |
| 7. | Application layer | Development of new models<br>Architecture | |

## Cognitive Radio Ad Hoc Networks

A *cognitive radio* detects the availability of unused licensed spectrum in its surroundings and dynamically configures itself to operate on that channel. The ISM band proposed for unlicensed used have been crowded due to emergence of plethora of computing devices that compete for access. The solution is to opportunistically capitalize on the licensed spectrums that are not optimally used in most of the cases. A *cognitive radio ad hoc network (CRAHN)* is an ad hoc network created among a set of nodes equipped with cognitive radios. There are two types of users in CRAHN. The *primary user* is the owner of a particular spectrum while the *secondary user* utilizes the unused spectrum of other user.

### Properties of CRAHN

CRAHN is differentiated from ad hoc networks by various properties (Akyildiz and Lee 2009). The nodes in CRAHN are equipped with *cognitive radios* that continuously sense the unused licensed spectrum. These radios should be self-describing and adapts its parameters to operate on different channel. There are *multiple channels* available to a particular user for transmission. The availability of channel for particular user varies continuously depending upon the activities of primary user. Hence, the routing path between two nodes usually comprises *communication in different spectrum*. The various layers of protocol are usually *coupled with the spectrum sensing mechanism*. For example, an application layer should be able to discriminate between the temporary unavailability of a link due to primary user's activity from other reasons. Additionally, the *neighborhood detection* by means of beacon message requires transmission in all available channels, which is not always feasible in CRAHN.

### Research issues in CRAHN

CRAHN presents several new challenges besides the conventional ad hoc network's challenges and therefore asks for deliberation on several new research issues. Among them, there are a range of research issues related to spectrum management. A cognitive radio user should *sense the spectrum* for unoccupied spectrum and identify spectrum holes. This step involves determining the primary user based on energy of the received signals or observing specific features i.e. modulation type, symbol rate and pilot signals etc. (Cabric, Mishra et al. 2004). An Eigen value based primary user detection method has been proposed in (Liu, Guo et al. 2013). In this scheme, a virtual multi-antenna structure is formed using temporal smoothing technique. A covariance matrix is thus obtained whose maximum and minimum Eigen values are used to detect primary user. However, the problems with local spectrum sensing are channel conditions and time varying nature of channel etc.

Besides performing the spectrum sensing locally, cooperative approaches can be adopted in which different users work together to perform spectrum sensing. In this direction, a collaborative spectrum sensing based on evolutionary game theory has been proposed in (Sasirekha and Bapat 2012). An adaptation algorithm is proposed such that network utility is maximized even in case of dynamic changes in the network. Cacciapuoti, Akyildiz et al. (2012) have also proposed a correlated based spectrum sensing scheme. The proposed distributed approach recommends selection of non-correlated users for cooperation during spectrum sensing. For interested readers, a survey of various spectrum sensing algorithms can be seen in (Yucek and Arslan 2009).

Once spectrum holes are detected, one of the holes is selected for use by user. Generally, a *spectrum decision approach* is based on characterizing the spectrum, selecting the best spectrum based on this characterization and reconfiguring the communication protocol and hardware according to the radio environment and available QoS (Akyildiz, Lee et al. 2009). Most of the approaches proposed for spectrum decision making are cooperative in nature. Lee and Ian F. Akyildiz (2011) have proposed a spectrum decision approach for real-time applications that considers channel dynamics and application requirements for selection. A cooperative weighted decision approach based on SRN tracking has been proposed in (Canberk and Oktug 2012). The proposal employed a distributed weighted fusion scheme to combine the decisions of individual users to obtain a cooperative decision.

As there may be multiple users trying to access the spectrum, there should be some coordination to allow *sharing of the spectrum* among multiple secondary users. Game theory has been proposed to provide the equilibrium condition where multiple users compete for the available spectrum (Neel 2006). A consensus based approach to spectrum sharing has been proposed in (Hu and Ibnkahla 2012). The scheme performs spectrum sharing based on local information (spectrum or location sensors) and consensus feedback.

For fair utilization of the spectrum among the users, a *MAC protocol* is also required. Unlike conventional ad hoc networks, MAC protocols in CRAHN should be strongly coupled with spectrum sensing process. Among the various protocols for CRAHN, there are protocols based on random access, time slotting and hybrid approaches (Akyildiz, Lee et al. 2009). *Random access protocols* are variants of CSMA/CA protocol that are not dependent on time synchronization and can access the medium any time. In *time slotted approaches*, appropriate time slots are provided to user for transmission of data and control frames. This requires global synchronization of nodes at network level. In the *hybrid approaches*, the control frames are transmitted based on time slotting but data can be transmitted randomly without any synchronization. A survey of various MAC protocols along with further details of these protocols can also be seen in (Cormio and Chowdhury 2009).

Upon detection of primary user, the cognitive radio user must vacate the spectrum. A *spectrum hand off* is thus required that ensures seamless communication by switching the communication to a new spectrum. For this purpose, proactive as well as reactive approaches can be adopted (Feng, Caoa et al. 2012). In *proactive approach* to hand-off, the future activity is predicted and a new spectrum is determined before the failure of current link. In *reactive approach*, switching to a new spectrum is performed after the links failure has occurred. This approach causes significant delay in current transmission.

A common *control channel* is required to for exchange of control information pertaining to spectrum management. This dedicated channel thus enables the cognitive radio users to seamlessly operate and supports neighborhood discovery, coordination in spectrum sensing and exchange of local measurements etc. (Akyildiz, Lee et al. 2009). Among the various choices for control channel include dedicated licensed spectrum, ISM and UWM bands (Marinho and Monteiro 2012).

*Routing* in CRAHN is not a trivial task as it must consider different aspects of spectrum management for route computation. Ideally, a route should be selected with the minimum channel switches to transmit a message to the destination. Among the various routing protocols for CRAHN are those based on utilizing spectrum information and middleware based approaches etc. (Chowdhury 2009; Guan, Yu et al. 2010). A survey on various routing protocols for CRAHN has been provided in (Cesana, Cuomo et al. 2010).

New *transport protocols* are required due to fluctuations in channel availability and temporary disconnections. As we discussed earlier, the TCP protocol suffer from issues due to its congestion control algorithm. Hence, the transport protocols for CRAHN should be spectrum aware. It should be able to discriminate between route disruptions due to channel switching and other reasons i.e. node mobility or node/ link failure. In CRAHN, as the failure is virtual and the same route is restored after a new channel is selected for transmission. Hence, instead of stopping the transmission the packets should be sent at an optimal route that doesn't overwhelm the intermediate nodes. An example of transport protocol for CRAHN is TP-CRAHN (Chowdhury, Felice et al. 2009). It is based on feedback from the physical and link layer to discriminate various types of events in the network. It also adopts a network layer predictive mobility framework that sends the packet at optimal rate to a sender. TCP-CR also proposed a delayed congestion control scheme based on primary user detection in cognitive radio networks (Yang, Cho et al. 2012).

CRAHN suffers from several additional security threats besides the security problems of conventional ad hoc networks. Some of the threats are jamming of control channels, masquerading of primary or cognitive radio users, unauthorized use of spectrum, integrity of cognitive radio messages and the node itself etc. (Baldini, Sturman et al. 2012). Fadlullah, Nishiyama et al. (2013) have presented an intrusion detection system for CRAHN that identifies various types of anomalies based on a learning model. A survey of various security challenges and solutions for cognitive radio networks have also been provided in (Attar, H. Tang et al. 2012).

An important issue is the development of new *applications models* for CRAHN. These application models should be agnostic against events happening at lower layers. For example, an application should be resilient against the temporary disconnection due to spectrum switches. New algorithms are required that focus on QoS at application layers. Yu, Sun et al. (2010) proposed an approach towards optimizing application layer QoS parameters using finite state Markov models. Finally, as the experimental evaluation of CRAHN on real environment is very complex, there needs to be *test beds and simulation toolkits* to validate various research proposals. Among the various software platforms and test beds proposed for CRAHN include GNU Radio and SCA (Ishizu 2006; Gonzalez 2009) etc. A survey on various implementation choices for CRAHN has been provided in (Chowdhury and Melodia 2010).

Table 6 summarizes various issues of CRAHN. There have been different variants of CRAHN proposed in literature that are evolved as intersection with other technologies. This section ends the discussion with an overview of them.

*Table 6: Summary of research issues in cognitive radio ad hoc network*

| | Research issue | Challenges / Issues | Representative approaches |
|---|---|---|---|
| 1. | Spectrum sensing | Sensing accuracy<br>Channel condition i.e. multipath effects, noise and interference etc.<br>Time varying signal properties | Cooperative approaches<br>Non-cooperative approaches |
| 2 | Spectrum decision | Power control<br>Time varying signal properties | (Canberk and Oktug 2012)<br>(Lee and Ian F. Akyildiz 2011) |
| 3 | Spectrum sharing | Energy management<br>Novel coordination mechanisms | Game theory (Neel 2006)<br>Consensus based spectrum sharing (Hu and Ibnkahla 2012) |
| 4. | MAC protocol | Coupling with spectrum sensing | Random access<br>Time slotting<br>Opportunistic access |
| 5. | Hand-off | Delay<br>Connection management | Proactive hand-off<br>Reactive hand-off |
| 6. | Control channel | Control channel contention<br>Jamming attack | Dedicated licensed spectrum<br>ISM band<br>UWM band |
| 7. | Routing | Primary user activity<br>New metrics considering spectrum awareness, channel properties, queuing and switching delay etc.<br>Mechanism for route maintenance | Utilizing spectrum information (Chowdhury 2009)<br>Middleware based approaches (Guan, Yu et al. 2010) |
| 8. | Transport protocol | Spectrum awareness | TP-CRAHN (Chowdhury, Felice et al. 2009)<br>TCP-CR (Yang, Cho et al. 2012) |
| 9. | Security | Jamming<br>Masquerading<br>Authorization<br>Integrity | (Fadlullah, Nishiyama et al. 2013) |
| 10. | Application | Spectrum agnostic<br>Novel Application models<br>QoS | (Yu, Sun et al. 2010) |
| 11. | Implementation | Performance<br>Accuracy | GNU Radio<br>SCA |

### Cognitive Radio Sensor Network (CRSN)

Akan, Karli et al. (2009) presented the applications, design and architecture of CRSN. The nodes in sensor networks can be configured in ad hoc style, arranged in the form of clusters or hierarchical, and in some cases, few nodes can be mobile. CRSN are especially useful in places where unlicensed spectrum band (ISM) is not available or there are many users competing for the particular spectrum. This includes applications in the domain of ubiquitous computing and ambient intelligence etc. CRSN additionally present various research issues i.e. power control, optimal node deployment, spectrum aware grouping of sensors, and analysis of optimal network coverage etc. (Akan, Karli et al. 2009; Nguyen and Hwang 2012).

### Cognitive radio vehicles (CRV)

Cognitive radio vehicles (CRV) are vehicular networks in which nodes are equipped with radios. They are useful for inter-vehicle communication, public safety and entertainment applications etc. Felice, Doost-Mohammady et al. (2012) have provided a summary of different research proposals on cognitive radio vehicles. CRV presents various new issues besides the issues associated with CRAHN. These are related to analyzing impact of high mobility vehicles on spectrum management, security and privacy issues, and development of new testbeds and toolkits for validation of research proposals. Representative solutions can be seen in (Felice, Chowdhury et al. 2011; Rawat, Zhao et al. 2013).

## FUTURE RESEARCH DIRECTIONS

We believe that the development on ad hoc networks and related technologies will eventually lead towards pervasive computing environment, a vision perceived by Mark Weiser (1995). Pervasive computing enables the assimilation of information processing in human's life in seamless way. The emergence of 4G wireless networks and corresponding technological developments are already paving the way. Ad hoc networks will play a key role in realization of this vision by enabling spontaneous erection of networks. These networks have the essential ingredients of self-organization, self-management and self-healing required for operations in ubiquitous environment. However, a lot of research efforts are still required before the true realization of pervasive phenomenon. Following paragraphs outline opportunities for further research in ad hoc networks.

For antenna design, current research is being done on design of directional and low dimension antennas. However, the impact of antenna on upper layers has to be analyzed. Research efforts are also required on devising low dimension smart antennas for its usage in practical applications. MAC layer protocols also require several modifications. Future research should be focused on energy efficient, multi-channel and cooperative MAC protocols and medium access control in presence of directional antennas. Since, ad hoc networks are zero-configuration systems; various research efforts are to be done on devising secure and scalable approaches for address assignments. The various routing protocols must consider the power constraints, the characteristics of channels along the path and other routing metrics to improve the routing operation. At the transport layer, not much work has been done. Protocols are required that utilize network events information from layers below for congestion and flow control. Security is also a challenging research area and research is required on addressing various associated challenges. Existing security solutions can't counter all types of security attacks as almost all of them are developed for a particular type of security threat. Generalized solutions are therefore required that can cope with any type of security threats. Ad hoc networks require cooperation enforcement mechanisms to enable distributed operation of various protocols. In this direction, various efforts are being put in to design error-free incentive based mechanisms to ensure node's cooperation. Data management has recently emerged as a hot topic of research. Several frameworks have recently emerged. However, most of the frameworks are still immature. Future work is required on designing a complete framework encompassing solutions for its various issues i.e. discovery, knowledge management, semantic data representation, and consistency management etc. Simulators and testbeds are vital tools for validation of any research proposal specifically for ad hoc networks. Different proposals have evolved in recent years. The problem with current simulator is that results obtained from a particular simulator are usually not similar when the same algorithm is executed on a different simulator. Research efforts are therefore required on accuracy and reliability of simulators. In addition, research should also be done on designing realistic mobility models that mimic real world situations. Future proposals should also exploit cross layer operations for energy efficient and QoS based MAC, routing and security solutions for ad hoc networks. Finally, the standardization of ad hoc network is an area that also requires attention.

Various models of ad hoc networks are emerging as discussed in previous sections. These new models provide various innovative applications however their potentials are still to be explored. At the same time, several additional challenges require attention from researcher's community. For example, ad hoc grid requires novel resource discovery and scheduling algorithms keeping into consideration dynamically changing QoS requirements of ad hoc environment. In addition, new models for security and economics are required to ensure cooperation among nodes. Similarly, ad hoc cloud requires new approaches for ensuring service provisioning, market oriented management of resources, data interoperability and privacy etc. Research efforts are required in improving data transmission, and addressing the issues arising in multi-channel MAC, routing, transport protocols and security of WMN. For CRAHN, the research is at initial stages and extensive efforts are required in development of techniques that can address spectrum management, mobility management, signalling, routing, transport protocol and security issues. Finally, new application models, implementation tools, analytical models and benchmarks have to be developed to assess the promise of all of these emerging models of ad hoc networks.

## CONCLUSION

This chapter discussed the research trends and issues in wireless ad hoc networks. It discussed in detail ad hoc networks and its various types. An overview of various challenges and issues of ad hoc networks have been presented and analysis of research efforts to address these issues has also been provided. We also discussed some of the most recent models of ad hoc network. The last part of this chapter presented several directions for further research.

## ADDITIONAL READING SECTION

## KEY TERMS & DEFINITIONS

Ad hoc network: A network that is formed on the fly without any prior planning and infrastructure

Cloud Computing: A collection of integrated resources that are dynamically provisioned as services over internet

Cognitive radio ad hoc network: An ad hoc network where the nodes are equipped with cognitive radios to use unused licensed spectrum in their vicinity

Grid Computing: An integration of loosely coupled resources belonging to different administrative domains for high performance computing

Mobile ad hoc network: An ad hoc network created among a set of mobile hosts

Sensor network: An ad hoc network established among a set of static hosts equipped with sensing capabilities

Vehicular ad hoc network: An ad hoc network formed between vehicles travelling on the road

Wireless mesh network: A type of network where a set of nodes autonomously organize themselves in the form of mesh topology to provide network access to different clients